\documentclass[structabstract]{aa}  
\usepackage{graphicx}
\usepackage{txfonts}
\usepackage[authoryear]{natbib}
\bibpunct[; ]{(}{)}{,}{a}{}{;}
\usepackage{longtable,lscape,rotating}
\begin{document}
   \title{An obscured cluster associated with the \ion{H}{ii} region RCW173\fnmsep\thanks{Partially based on observations  
collected at the Nordic Optical Telescope (La Palma)}}

    \author{A. Marco\inst{1}
          \and
          I. Negueruela\inst{1}
          }


   \institute{Departamento de F\'{i}sica, Ingenier\'{i}a de Sistemas y Teor\'{i}a de la Se\~{n}al. Escuela Polit\'ecnica Superior. University of Alicante. Apdo.99 E-03080. Alicante. (Spain)\\
              \email{amparo.marco@ua.es}}

   \date{Received; accepted}
   \titlerunning{Alicante~6 and RCW~173}

 
  \abstract
   {The discovery of several clusters of red supergiants towards $l=24\degr$--$30\degr$ has triggered interest in this area of the Galactic plane, where lines of sight are very complex and previous explorations of the stellar content were very preliminary.}
   {We attempt to characterise the stellar population associated with the \ion{H}{ii} region RCW~173 (=Sh2-60), located at $l=25\fdg3$, as previous studies have suggested that this population could be beyond the Sagittarius arm.}
   {We obtained $UBV$ photometry of a stellar field to the south of the brightest part of RCW~173, as well as spectroscopy of about twenty stars in the area. We combined our new data with archival 2MASS near-infrared photometry and {\it Spitzer}/GLIMPSE imaging and photometry, to achieve more accurate characterisation of the stellar sources and the associated cloud.}
   {We find a significant population of early-type stars located at $d=3.0$~kpc, in good agreement with the ``near'' dynamical distance to the \ion{H}{ii} region. This population should be located at the near intersection of the Scutum-Crux arm. A luminous O7\,II star is likely to be the main source of ionisation. Many stars are concentrated around the bright nebulosity, where GLIMPSE images in the mid infrared show the presence of a bubble of excited material surrounding a cavity that coincides spatially with a number of B0-1\,V stars. We interpret this as an emerging cluster, perhaps triggered by the nearby O7\,II star. We also find a number of B-type giants. Some of them are located at approximately the same distance, and may be part of an older population in the same area, characterised by much lower reddening. A few have shorter distance moduli and are likely to be located in the Sagittarius arm.}
   {The line of sight in this direction is very complex. Optically visible tracers delineate two spiral arms, but seem to be absent beyond $d\approx3$~kpc. Several \ion{H}{ii} regions in this area suggest that the Scutum-Crux arm contains thick clouds actively forming stars. All these populations are projected on top of the major stellar complex signposted by the clusters of red supergiants.}


   \maketitle
%

\section{Introduction}

In the past few years, several clusters of red supergiants have been discovered in a small region
of the Galactic plane, between $l=24\degr$ and $l=29\degr$ (e.g., \citealt{davies2007}; \citealt{clark2009a}).
These clusters are very distant ($d\approx6$~kpc) and lie behind heavy obscuration \citep[e.g.,][]{negueruela11}, implying that only the bright red supergiants have been
detected so far. In spite of this, the clusters are believed to be very massive ($M_{\rm {cluster}}\approx 2-5\times10^{4}M_{\sun}$), because population synthesis models predict that red supergiants are very
rare \citep{clark2009b}. Several authors have suggested the existence of a giant star-forming region
associated with the intersection of the base of the Scutum-Crux arm and the long Galactic bar \citep[e.g.,][]{garzon1997,davies2008}, though the connection between the different clusters is still unclear \citep{negueruela10a}. Continued star formation in this region is likely according to the analyses of radio and infrared data towards $l=30\fdg5$, which find evidence of a very large star-forming region around W43, also at $d\approx6$ kpc \citep{bally10,nguyen11}.

The distribution of stars and dust in this region is poorly known. Clusters in this direction are difficult to study because of background confusion. There appear to be a number of clusters at the distance of the Sagittarius arm, none of which
is very young. The most well-studied are NGC~6664 ($l=24\degr$) and Trumpler~35 ($l=28\fdg3$), both at distances $\la2$ kpc and moderately reddened, e.g., $E(B-V)\approx1.0$ for Trumpler~35 \citep{turner1980}.  

We are carrying out detailed studies of cluster candidates possibly associated with the putative cluster complex at $\sim6$~kpc \citep[e.g.,][]{negueruela10a,negueruela11} to derive a census of star formation activity. In an attempt to find other distance calibrators in this very interesting area, we explore the \ion{H}{ii} region 
RCW~173 ($l=25\fdg3$, $b=+0\fdg3$) \citep{rodgers60}, also known as Sh2-60 \citep{sharpless59} or LBN~$025.36+00.24$. Previous authors had associated this bright nebulosity with moderately bright early-type stars.  

\citet{roslund1963} carried out a photographic survey of early-type stars covering seven square degrees around RA(J1950) 18h 37m and Dec(J1950) $-07\degr 12\arcmin$ (18h 39m 42s, -07\degr 9\arcmin 13\arcsec in J(2000). The purpose of this survey was to study the absorption and space distribution of B stars in the area. As a result of this catalogue, six early-type stars were identified in the vicinity of RCW 173. From their photometry, \citet{roslund1963} assumed that all those stars were mid and late B main-sequence objects. \citet{vogt75}, instead, studied the possibility that two of those stars might be the ionisation source of RCW 173, concluding that [R63]~34 = GSC~05123-02611 (our star a803) could be the source of ionisation, based on its photometric properties (probably assuming that it was a main-sequence O type star instead of a B1\,II giant, as we find later). Finally, \citet{lahulla85} observed
these two stars again, plus a third object that seems to be the blend of several faint stars immersed in nebulosity, concluding that they were all early-type stars.

In this paper, we present a spectroscopic and photometric study of the area surrounding RCW 173. We find several populations of early-type stars at different distances and reddenings. The most heavily reddened population corresponds to an open cluster emerging from the cloud associated with the \ion{H}{ii} region RCW 173. We call this new cluster Alicante 6.

In section 2, we present our observations. In section 3, we describe the photometric and spectroscopic analysis, while in section 4 we discuss the implications of our findings to understand this complex sightline. 


\section{Observations and data}

We obtained $UBV$ photometry of the region around RCW 173 and spectroscopy of selected stars in the field using ALFOSC on the Nordic Optical Telescope at the Roque de los Muchachos Observatory (La Palma, Spain) on the nights of 27-29 June 2009. Spectroscopy was taken over the three nights, while the photometry was taken during the last night, which was the only one with photometric conditions.

\subsection{Photometry}

\begin{figure*}[ht]
\resizebox{10 cm}{!}{\includegraphics{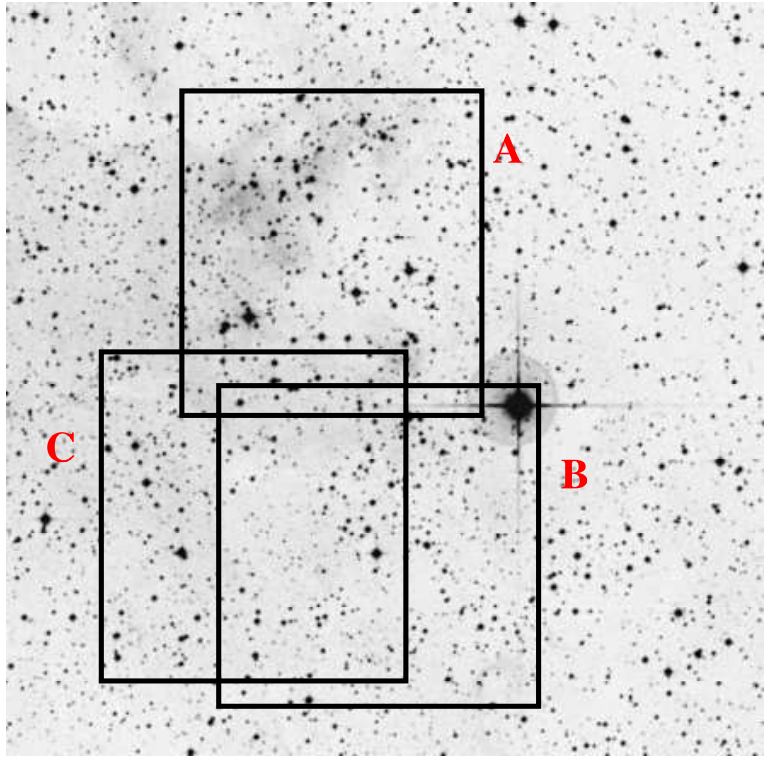}}
\centering
\caption{Map of the area observed in the field of RCW 173. The central coordinates of each frame are given in Table~\ref{t1}. $XY$
  positions are listed in Table~\ref{t2} for stars in each frame. The origin of coordinates is located at
  the bottom left corner of each frame. North is up and east is left.} 
\label{Fig1}
\end{figure*}

ALFOSC allows observations in different modes. In imaging mode, the camera covers a field of $6\farcm5 \times 6\farcm5$ and has a pixel scale of $0\farcs19/$ pixel. Since the area of interest is wider than the field of view, we took three partially overlapping frames (named A,B, and C in Fig.~\ref{Fig1}), covering an area of approximately $8\farcm0 \times 13\farcm0$, which includes the brightest nebulosity in RCW 173 and the blue stars to the south. The area to the north of the H\,{\sc ii} region was not observed because the DSS2 images suggest that the dark cloud covers this area. For each frame, we obtained three series of different exposure times in each filter to achieve accurate photometry for a broad magnitude range. The central positions of each frame and the exposure times used are presented in Table~\ref{t1}. The whole field is shown in  Fig.~\ref{Fig1}. 

Fifteen standard stars belonging to the SA~110 field from the list of \citet{landolt}
were observed several times during the night to trace
extinction and provide standard stars for the transformation. Their
images were processed for bias and flat-fielding corrections with the
standard procedures using the CCDPROC package in IRAF\footnote{IRAF is
  distributed by the National Optical Astronomy Observatories, which
  are operated by the Association of Universities for Research in
  Astronomy, Inc., under cooperative agreement with the National
  Science Foundation.}. Aperture photometry using the PHOT package
inside DAOPHOT (IRAF, DAOPHOT) was developed on these fields with
the same aperture, 15 pixels, for each filter.

\begin{table}
\caption{Log of the photometric observations taken at the NOT on June 2009 for RCW 173. Typical image quality is approximately $1.0\arcsec$}
\label{t1}
\centering
\begin{tabular}{c c c}
\hline\hline
Field A & RA(J2000) = 18h 36m 22.5s & Dec(J2000) = $-06\degr 41\arcmin 36\arcsec$ \\
Field B & RA(J2000) = 18h 36m 19.0s & Dec(J2000) = $-06\degr 47\arcmin 26\arcsec$ \\
Field C & RA(J2000) = 18h 36m 29.2s & Dec(J2000) = $-06\degr 47\arcmin 00\arcsec$ \\
\end{tabular}
\begin{tabular}{c c c c}
\hline\hline
&\multicolumn{3}{c}{Exposure times (s)}\\
Filter & Long times & Intermediate times & Short times \\
\hline
U & 900 & 200 & 15 \\
B & 250 & 40 & 4 \\
V & 120 & 20 & 2 \\
\hline
\end{tabular}
\end{table}

The reduction of the frames obtained in fields A,B and C around RCW 173 was done with IRAF routines
for the bias and flat-field corrections. Photometry was obtained by
point-spread function (PSF) fitting using the DAOPHOT package
\citep{stetson1987} provided by IRAF. To construct the PSF empirically,
we automatically selected bright stars (typically 25 stars). After this,
we reviewed the candidates and discarded those that do not reach the
optimal conditions for a good PSF star according to \citet{stetson1987}. We insisted that, a good PSF star was not blended, nor had close neighbours of similar brightness, nor was affected by bad pixels. Once we had a list of PSF
stars ($\approx 20$), we determined an initial PSF by fitting the best-fit
function between the five options offered by the PSF routine inside
DAOPHOT. We allowed the PSF to be  variable (to order two) across the
frame to take into account the systematic pattern of PSF variability
with position on the chip. 

We needed to perform aperture correction for each frame in all filters.
Finally, we obtained the instrumental magnitudes for all stars. Using
the standard stars, we carried out the atmospheric extinction
correction and transformed the instrumental magnitudes to the
standard system using the PHOTCAL package inside IRAF.  

We complemented our dataset with $JHK_{s}$ photometry from the 2MASS
catalogue \citep{Skrutskie2006}.

The number of stars that we could detect in all filters is limited by
the long exposure time in the $U$ filter. In
Table~\ref{t2}, \ref{t3}, and~\ref{t4}, we list their $X$ and $Y$
positions with respect to each frame shown in Fig.~\ref{Fig1}, and their identification with objects in
the 2MASS catalogue, together with their coordinates (right
ascension (RA) and declination (DEC) in J2000). A few stars have no
obvious corresponding source in the 2MASS catalogue. We identified
these objects in the GLIMPSE catalogue \citep{glimpse} and assume their RA and DEC
coordinates to be those assigned in this catalogue. A few stars have no counterparts in any of the two catalogues. The designation of each
star is given by the letter that identifies the frame on which it is detected, plus a sequential number. Stars observed in more than one frame receive the designation corresponding to the frame with the highest number of measurements. 

We have photometry for roughly 600 stars in the fields. In
Table~\ref{t2}, \ref{t3}, and~\ref{t4}, we list the values of $V$,
$(B-V)$, and $(U-B)$ with the standard deviation and
the number of measurements for each magnitude and index.

\subsection{Spectroscopy}  

In spectroscopic mode, we used grisms \#14 and \#16 combined with a $1\arcsec$ slit to obtain intermediate resolution spectroscopy. Grism \#14 covers the 3275--6125 \AA\ range with a nominal dispersion 1.4 \AA/pixel. Grism \#16 covers the 3500--5060 \AA\ range with a nominal dispersion 0.8 \AA/pixel. We observed the six stars given by \citet{roslund1963} as B-type stars, plus a number of other objects which we selected as likely early-type stars based on their positions on the 2MASS CMDs (following the procedures discussed in \citealt{cp05} and \citealt{ns07}). In total, we observed 18 stars, which are listed in Tables 2 and 3. 

All spectroscopic data were reduced using the Starlink software
packages CCDPACK \citep{Draper2000} and FIGARO \citep{Shortridge1997}. We used standard procedures for bias subtraction and flat-fielding (using internal halogen lamps). Wavelength calibration was achieved by using ThAr arc lamp spectra, taken between the exposures. The rms for the wavelength solution is $\approx0.2$ pixels for both grisms. The spectra were normalised to the best-fit function of the continuum using DIPSO \citep{Howarth1998}.

\section{Results}

\subsection{Spectroscopy}

We used the spectra obtained for spectral classification. Classification was made by comparison to spectra of MK standard stars observed at similar resolution, and to the digital atlas of \citet{walborn1990}. We checked internal consistency by comparing all the spectra obtained amongst themselves. This procedure allows for accurate spectral classification. In practice, the accuracy of the classification is limited by the low signal-to-noise ratio (SNR) of some spectra in the 4000--4500 \AA \ region.

\begin{figure}
   \centering
   \resizebox{\columnwidth}{!}{\includegraphics[angle=-90]{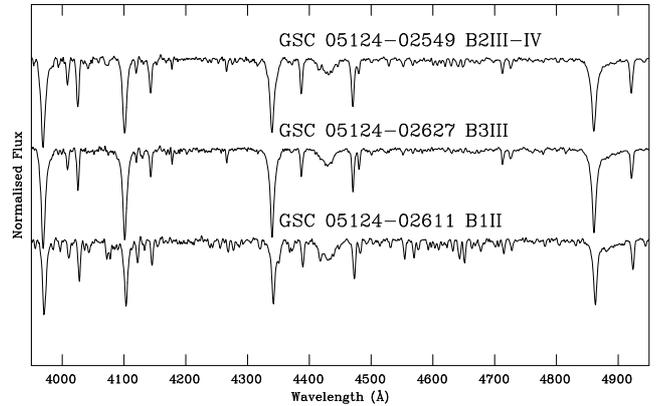}}
   \caption{Classification spectra of three stars included in the work of \citet{roslund1963}, which turn out to be early B giants. GSC~05124-02611 (a802) is clearly more luminous than the other two, and close to the supergiant classification.
           }
   \label{fig:bgiants}
    \end{figure}

\begin{figure}
   \centering
   \resizebox{\columnwidth}{!}{\includegraphics[angle=-90]{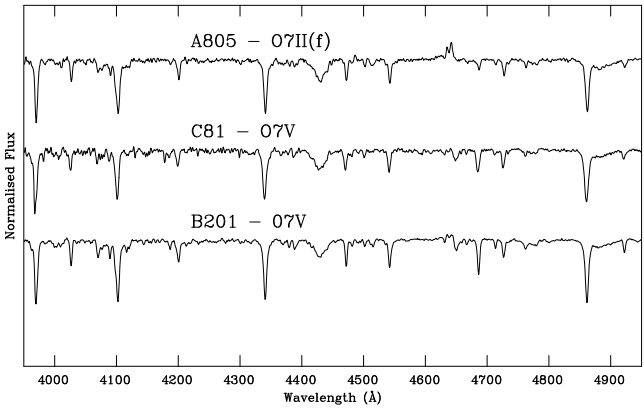}}
   \caption{Classification spectra of the three O-type stars found in the field. Star a805 shows signs of high luminosity, such as the strong \ion{N}{iii} emission and filled-in \ion{He}{ii}~4686\AA. This object is close to being a supergiant. C81 and B201 have lower luminosities. There is some evidence of (at least) a second, later-type star in the spectrum of C81, but this cannot be ascertained with the existing spectrum. 
           }
   \label{fig:ostars}
    \end{figure}

 The spectral types of stars observed with grism \#16 can be considered accurate to  $\pm0.5$ spectral types, which is the typical accuracy of the procedure. Grism \#14 was mainly used for fainter stars. These spectra have lower resolution and, because of higher extinction, lower SNR in the blue, and we expect them to have an accuracy of $\pm1$ spectral types. Even though most of the stars observed with this grism have spectral types in the B0--B1 range, where the MK grid is thick, the low SNR and resolution do not permit a good separation of the weak metallic lines (mostly \ion{Si}, but also \ion{O}{ii}) defining the spectral subtypes. Some of the spectra are displayed in Figures~\ref{fig:bgiants}, \ref{fig:ostars}, and~\ref{fig:oldstars}.  

We find that the stars identified by \citet{roslund1963} are 
a collection of late B type stars and early B giants. Amongst fainter (in the optical) stars, we find three O type objects and a large number of B0--1 main sequence objects. The list of objects for which we have photometry and spectroscopy is given in Table~\ref{t5}. The spectral types of other objects outside the area that we cover photometrically are given in Table~\ref{t6}. 

\begin{table*}
\begin{minipage}[t]{\textwidth}
\caption{Spectral types and photometry for stars in the
  region. Stars marked with ``*'' have been observed with grism \#14. Stars not marked have been observed with grism \#16. The intrinsic colours from \citet{fitzgerald1970} are used to derive excesses and dereddened magnitudes. These agree well with those obtained from CHORIZOS, which should be considered more accurate, as the code takes colour terms into full account.\label{t5}}
\centering
\begin{tabular}{lccccccccc}
\hline\hline
Star&Spectral Type&$V$&$B-V$&$U-B$&$(B-V)_{0}$&$(U-B)_{0}$&$E(B-V)$&$E(U-B)$&$V_{0}$\\
\hline
a801$^{*}$	&B1\, V	   &12.559&0.682&-0.171&-0.260&-0.950&0.942&0.779&9.639\\
a802	&B1\, II   &11.525&1.211&0.101&-0.240&-0.990&1.451&1.091&7.027\\
a803	&B2\, III-IV&12.152&0.712&-0.068&-0.240&-0.930&0.952&0.863&9.201\\
a805	&O7\, II	&12.827&1.833&0.446&-0.320&-1.170&2.153&1.616&6.153\\
A507$^{*}$	&B1\, V	&14.733&1.554&0.373&-0.260&-0.950&1.814&1.323&9.110\\
A263$^{*}$	&B1\, V	&14.719&1.546&0.360&-0.260&-0.950&1.806&1.310&9.120\\
A253$^{*}$	&B0.5\,V&14.646&1.582&0.406&-0.280&-1.000&1.862&1.406&8.874\\
A197$^{*}$	&B1\, V	&14.900&1.630&0.504&-0.260&-0.950&1.890&1.454&9.041\\
A514$^{*}$	&B0.5\, V&14.674&1.923&0.641&-0.280&-1.000&2.203&1.641&7.845\\
A515$^{*}$	&B0.5\, V&14.414&1.881&0.624&-0.280&-1.000&2.161&1.624&7.715\\
b201	&O7\, V	&12.434&1.261&0.059&-0.320&-1.170&1.581&1.229&7.533\\
C81	&O7\, V	&12.605&1.726&0.390&-0.320&-1.170&2.046&1.560&6.262\\
C84	&B1\, V	&12.636&0.863&-0.075&-0.260&-0.950&1.123&0.876&9.155\\
C129	&B7\, III&13.025&0.785&0.235&-0.120&-0.440&0.905&0.675&10.220\\
\hline
\end{tabular}
\end{minipage}
\end{table*}

\begin{table*}
\begin{minipage}[t]{\textwidth}
\caption{Spectral types and photometry for stars in the
  region not covered by our photometry. The three GSC stars were observed with grism \#16. $UBV$ photometry is from \citet{roslund1963}. The intrinsic colours from \citet{fitzgerald1970} are used to derive excesses and dereddened magnitudes. This is not attempted for the chemically peculiar star.\label{t6}
}
\begin{center}
\begin{tabular}{lccccccccc}
\hline\hline
Star&Spectral Type&$V$&$B-V$&$U-B$&$(B-V)_{0}$&$(U-B)_{0}$&$E(B-V)$&$E(U-B)$&$V_{0}$\\
\hline
GSC 05124-02627&B3\, III&11.82&0.95&0.17&-0.20&-0.73&1.15&0.90&8.26\\

GSC 05124-02567&B9\, IV   &12.04&0.76&0.50&-0.07&-0.19&0.83&0.69&9.47\\

GSC 05124-02605&B6\, III-IVp Si&12.66&0.85&0.25&-&-&-&-&-\\

2MASS~J18363487-0637227\tablefootmark{1}	&B0.5\, V&-&-&-&-&-&-&-&-\\
 
\hline
\end{tabular}
\newline\\
\end{center}
\tablefoottext{1}{This star lies $3\farcm5$ NE of Alicante~6. It has $(J-K_{\rm S})= 0.70$ implying $E(J-K_{\rm S})= 0.9$. The USNO catalogue gives $B1=14.7$ and $B2=15.5$. These values suggests that it is a cluster member }
\end{minipage}
\end{table*}

\subsection{Observational HR diagram}

\begin{figure}
   \resizebox{\columnwidth}{!}
   {\includegraphics{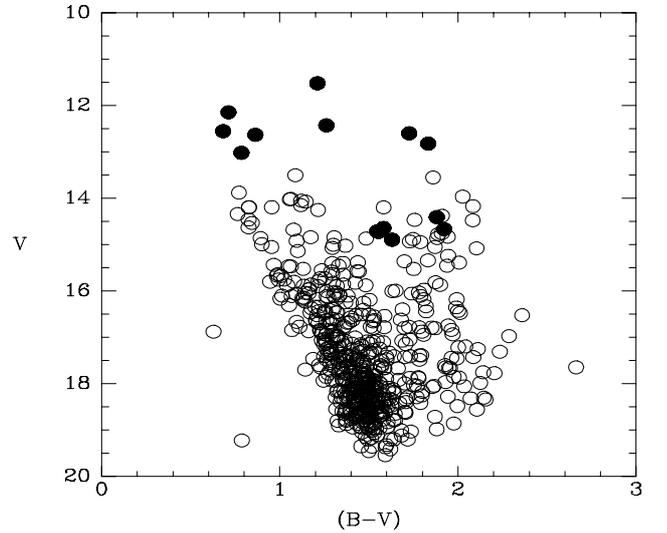}}
   \caption{$V/(B-V)$ diagram for all 
            stars in the field around RCW 173. The filled circles represent 
            early-type stars that have been spectroscopically observed.
           }
   \label{Fig3}
    \end{figure}

We start the photometric analysis by plotting the $V/(B-V)$ diagram
for all stars in our fields (see Fig.~\ref{Fig3}). The diagram does not show a clear sequence, probably because of strong
contamination by field stars and differential reddening. This circumstance forces us to follow  a
careful analysis procedure.

\subsubsection{The reddening law}

The first step is to determine whether the extinction law in the
direction of the field is standard. We use the CHORIZOS ($\chi^2$ code
for parametrised modelling and characterization of photometry and
spectroscopy) code developed by \citet{jesus2004}. This code fits
synthetic photometry derived from the 
spectral energy distribution of a stellar model convolved with 
an extinction law to reproduce the observed magnitudes.

We use as input to CHORIZOS a file with the $UBVJHK$ photometry and the 
$T_{{\rm eff}}$ corresponding to the spectral types 
derived according to the calibrations of \citet{fitzgerald1970}. The
output of CHORIZOS is the value of $R$ for each star. The average value for our sample is
$R=3.07\pm0.09$. This is taken as confirmation that the extinction law is standard in this direction.

\subsubsection{The reddening-free $Q$ parameter and spectral types}

The reddening-free $Q$ parameter allows a preliminary selection of
early-type stars. The $Q$ parameter is defined as

\begin{equation}
 Q=(U-B)-\frac{E(U-B)}{E(B-V)}(B-V)
\, .
\end{equation}

\begin{figure}
   \resizebox{\columnwidth}{!}
   {\includegraphics{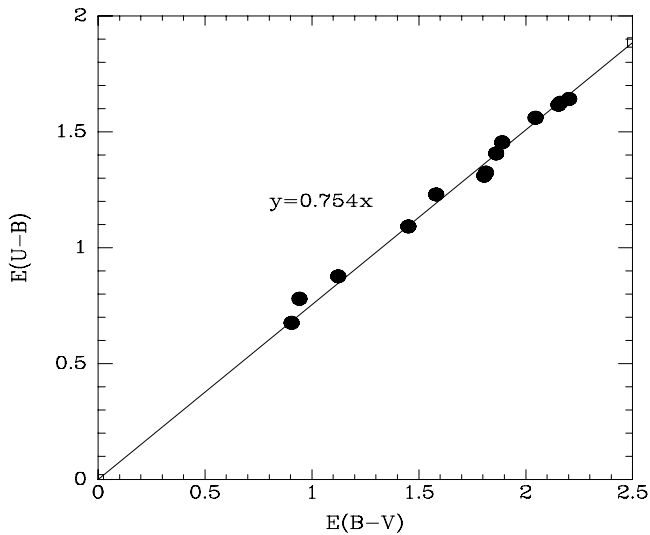}}
   \caption{$E(U-B)/E(B-V)$ diagram for stars with spectroscopy and photometry.
           }
   \label{Fig4}
    \end{figure}

For a standard reddening law,  $E(U-B)/E(B-V)=0.72$
\citep{johnson1952}. For the  
stars in our dataset with spectra and $UBV$ photometry, we calculate
the value of $E(U-B)$ and $E(B-V)$, taking  intrinsic colours from
\citet{fitzgerald1970}. In Fig.~\ref{Fig4}, we represent the 
$E(U-B)$/$E(B-V)$ diagram for these stars and we calculate the equation of the line
that best fits those points using the linear least squares fitting technique.

We obtain an average value of $0.75\pm0.04$ ($R^{2}=0.99$). The standard value for the slope, 0.72, is
compatible within the errors. However, when we calculated $Q$ values using the standard slope, we found that the $Q$ (photometric) spectral types were in most cases one subtype later than those derived from the spectra (for the stars
with spectroscopy). This suggests that, even though the standard value is within the errors, we must use the higher value $0.75$.  In nearby fields, \citet{turner1980} also found the slope to be consistently slightly steeper than the standard and close to 0.75. Given the high reddenings in our objects, even such a small difference may be important. Therefore we recalculated all the $Q$ values using 0.75.

   \begin{figure}
   \centering
   \includegraphics[width=\columnwidth]{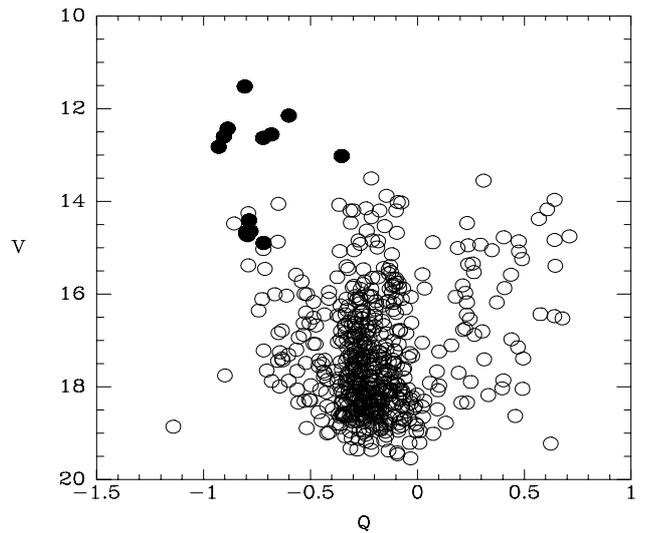}
   \caption{$V/Q$ diagram for stars in the field around RCW 173.The solid circles represent early-type stars spectroscopically observed.\label{qv}}
    \end{figure}

With these new $Q$ values, we plotted the $V/Q$ diagram displayed in Fig.~\ref{qv}, in which we can discern a sequence of early-type stars (defined by $Q<-0.5$) and a very important population of late B type stars (with $-0.5\leq Q\leq0$). The late B stars cover a vast range in $V$, indicating that they have different distances or reddenings. Indeed, most of the stars in the field have $Q$ between $-0.4$ and $-0.1$, which corresponds to spectral types between B6 and B9. Most stars with $Q\leq-0.5$ (i.e., earlier than B5) display a good correlation between $Q$ and V, suggesting that these are a single population of early-type stars. Many of these objects are concentrated towards the bright nebulosity on the northern edge of the field. The sequence, however, is broad, suggesting that reddening is variable. In view of this, confirming that these objects represent a single population will require a more detailed analysis.  We use the 
$(U-B)$/$(B-V)$ diagram for stars with $Q<0$ to derive approximate spectral type (this is equivalent to the $Q$ spectral typing).

 The $Q$ parameter was calibrated mostly using stars less reddened than those in the region. The calibration does not take into account bandpass effects, which may be important at high reddening. For this reason, the pseudo spectral types derived should only be considered approximations, which will be used as input to CHORIZOS. The code takes into account bandpass effects.
 
Using the approximate spectral types, we derive a preliminary estimate of the intrinsic $(B-V)_{0}$ by selecting the colour corresponding to the spectral type in the calibration of \citet{morton68} under the assumption that all stars are dwarfs. This allows a first estimation of the reddening.

We then deredden the $V$ magnitude using this approximation to obtain an estimate of the $V_{0}$ magnitude. Utilising these $V_{0}$ values for the stars earlier than B5, we make a first estimate of the average distance modulus of the population, obtaining $DM=12.4$. After this, we reject stars later than B5 with photometric distance moduli deviating by more than 1.5 magnitudes from this average. This procedure is should remove the B-type stars that are not associated with the O-type stars. For stars associated with the OB population, the assumption of a main-sequence nature is justified.

\subsubsection{Determination of the distance}

This refined selection procedure still leaves us with a large number of candidate members of the population associated with the O-type stars. In this sample, we again use the CHORIZOS code.  In this case, we use as input data the $UBVJHK_{\rm s}$
magnitudes (our photometry + 2MASS), the $T_{{\rm eff}}$ corresponding to
the photometric spectral types in the previous section
(according to the calibration of \citealt{fitzgerald1970}), and $\log g$ values
appropriate for a main-sequence star. The output from CHORIZOS are the
individual values of the excess $E(B-V)$ and the dereddened $V$
magnitudes. With the dereddened magnitudes and colours, we plot the $V_{0}/(B-V)_{0}$ diagram. We use the observational ZAMS from \citet{skaler82} shifted to different distance moduli to determine the distance to the cluster, confirming the value $DM=12.4$.

   \begin{figure}
   \centering
\resizebox{\columnwidth}{!}{\includegraphics[width=\columnwidth]{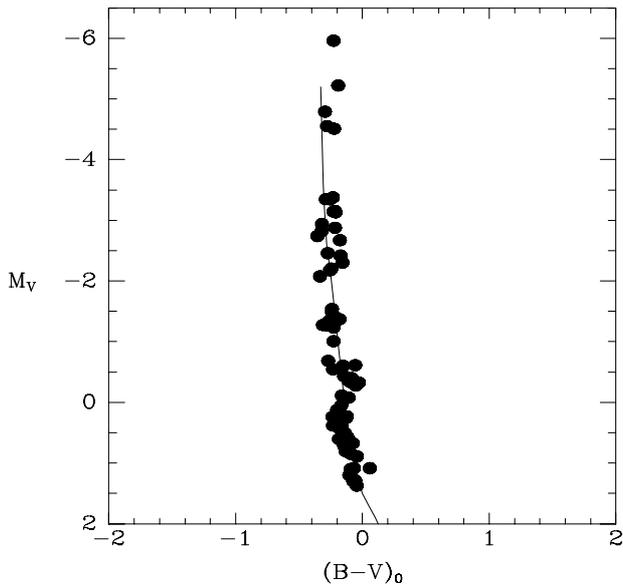}}
   \caption{$M_{V}/(B-V)_{0}$ diagram for very likely members. The thin line is the ZAMS from \citet{skaler82}  \label{CMDclean}}
 \end{figure}

After we determine $M_{V}=V_{0}-DM$ for each star, we plot the diagram shown in Fig.~\ref{CMDclean}.
The dereddened sequence gives a good fit to the ZAMS. However, because of very variable reddening and the obvious presence of contaminating populations (see also below), we assume a relatively conservative error of $\pm 0.3$ for our $DM$, and that $DM=12.4\pm0.3$, based on 74 very likely members. This $DM$ corresponds to a distance of $3.0^{+0.6}_{-0.4}$ kpc. This distance suggests that the cluster is located in the Scutum-Crux arm and definitely places this population well in front of the clusters of red supergiants. In Table~\ref{t7}, we show the dereddened values obtained by CHORIZOS for all the likely members.

\begin{figure*}[ht]
\resizebox{10 cm}{!}{\includegraphics{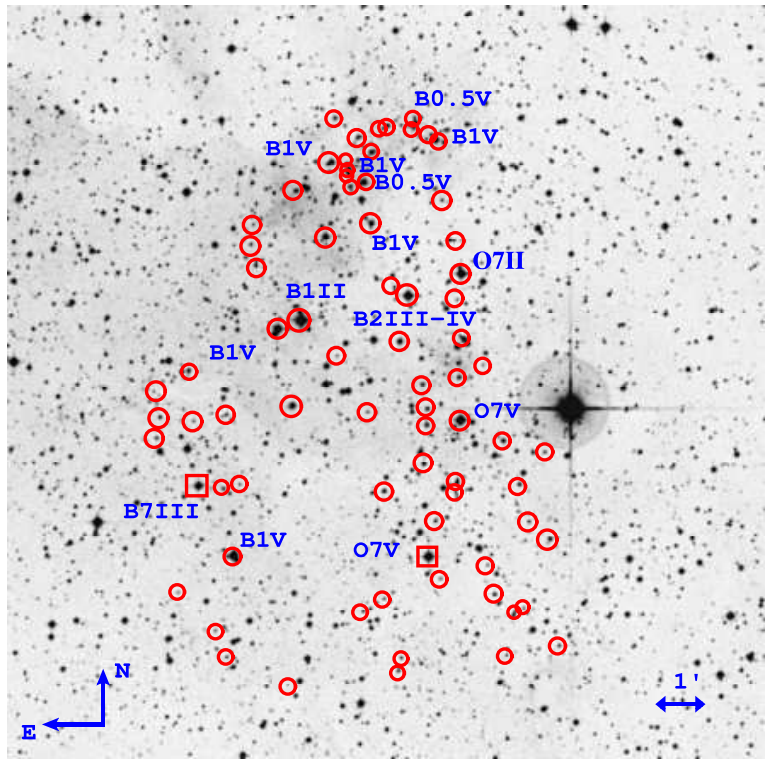}}
\centering
\caption{Map showing all the likely members of the new cluster Alicante 6 and the surrounding association that fits the ZAMS at $DM=12.4$ (circles). The two objects marked with squares have dereddened magnitudes incompatible with their spectral types. C81 (O7\,V) is more than one magnitude brighter than expected. Its spectrum suggests that we are seeing at least two unresolved stars. C129 (B7\,III), on the other hand, cannot be associated with Alicante 6 and is likely to be on the Sagittarius arm\label{mapmembers} } 
\end{figure*}

\section{Discussion}

We have characterised the population of early-type stars in the vicinity of the \ion{H}{ii} RCW 173. In spite of very strong differential reddening, most of the OB stars deredden to fit the ZAMS with $DM=12.4\pm0.3$. In particular, a subset of these objects, mostly those with the highest reddening, are concentrated strongly towards the bright nebulosity and very clearly seem to constitute a very young cluster in the process of emerging from its parental cloud.
We call this cluster Alicante 6. 

Figure~\ref{mapmembers} shows the distribution of the 74 members selected. The spatial distribution of the stars does not resemble a concentrated cluster.  In addition, when we consider the spectral types of those stars for which we have classification spectra, it is quite clear that not all the stars that fit the ZAMS at $DM=12.4\pm0.3$ are directly associated with the cluster. Some of the stars are early B giants, and therefore must belong to a population with age $\ga10$ Myr. We now attempt to clarify the connection between this population and the cluster.

\subsection{The very young cluster Alicante 6}

\begin{figure*}
\centering
\includegraphics[angle=0, width=\columnwidth]{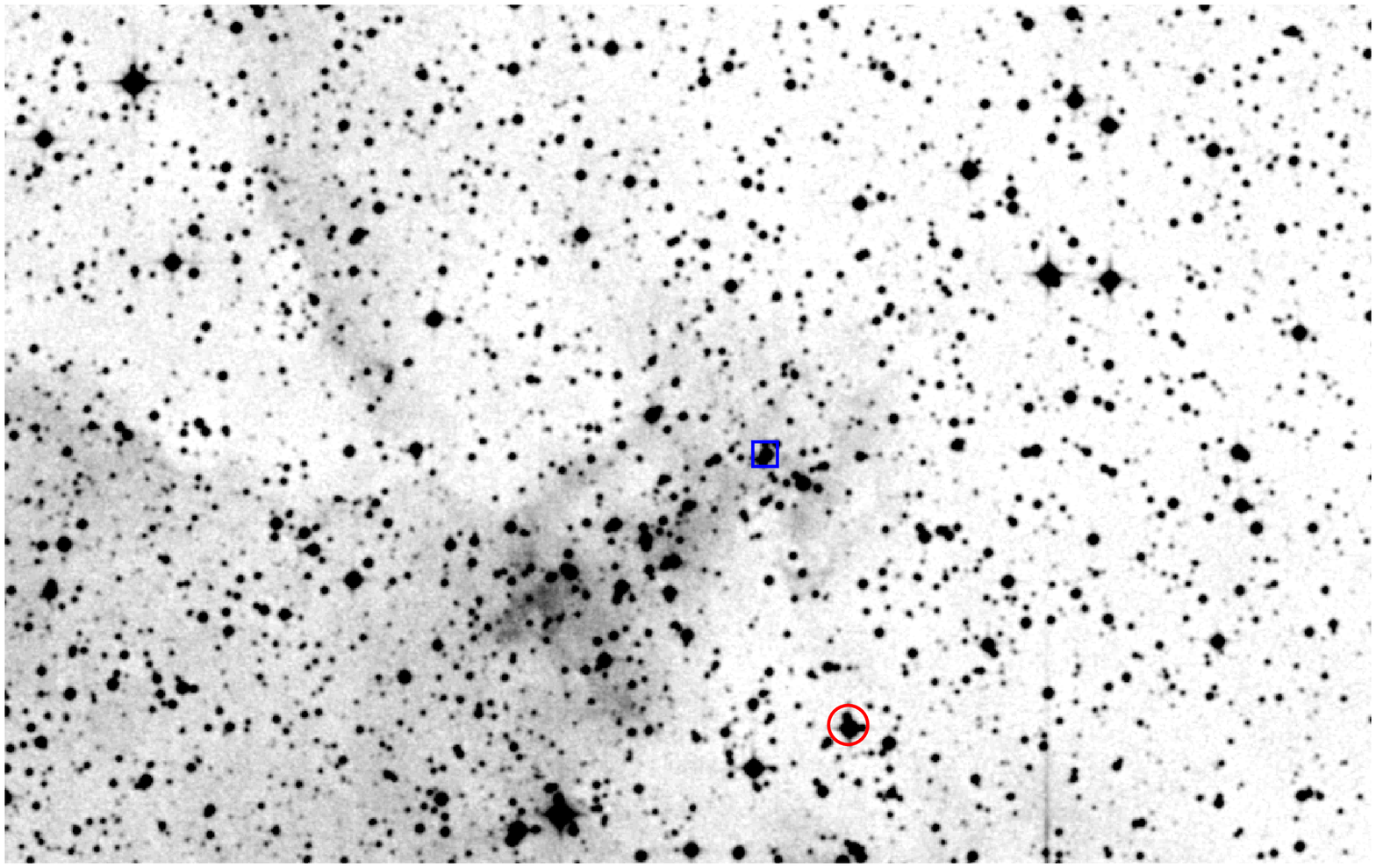}
\includegraphics[angle=0, width=\columnwidth]{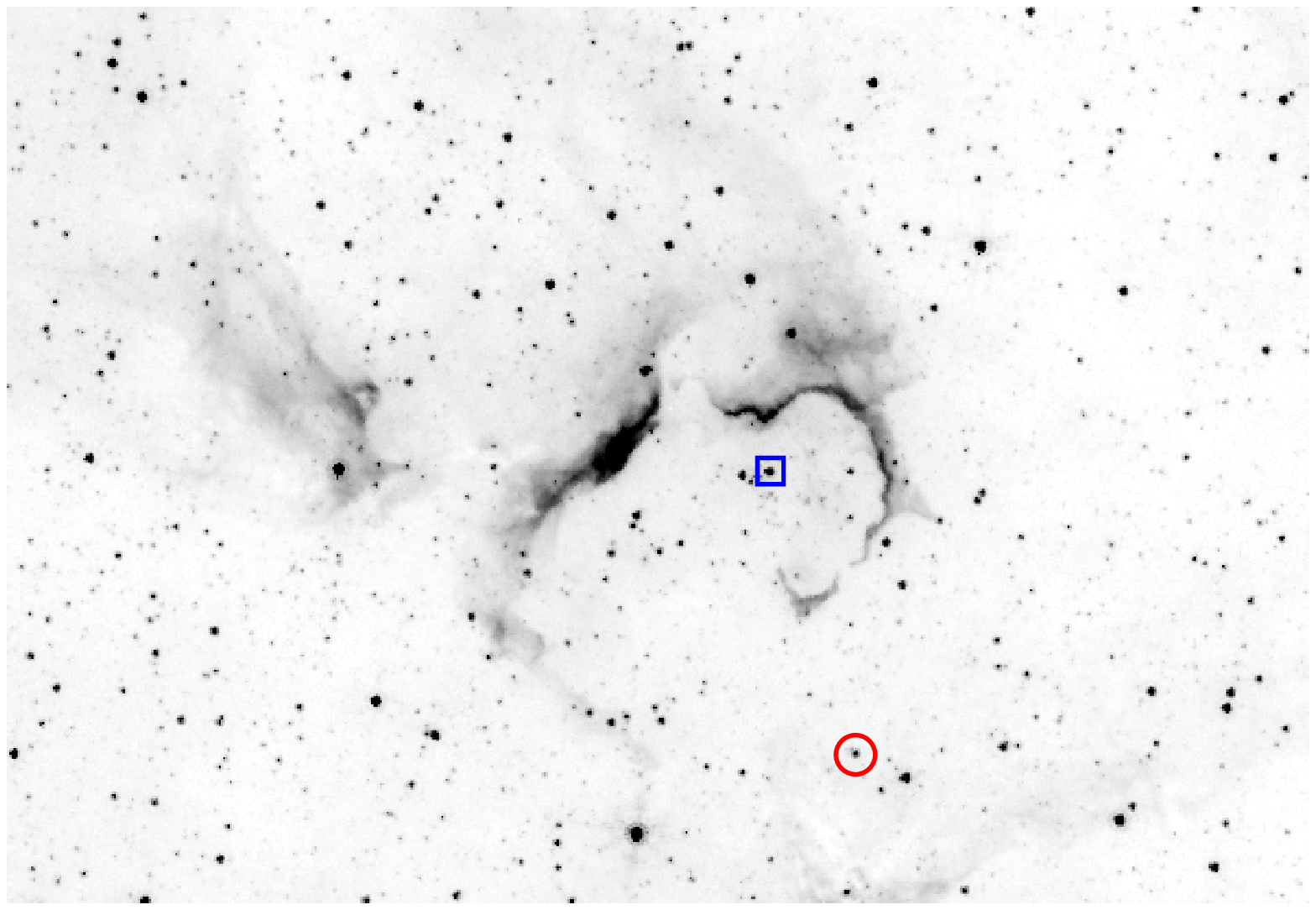}
\caption{{\bf Left panel: } DSS2 red image of the brightest nebulosity in RCW~173, and the emerging cluster Alicante~6. {\bf Right panel: } {\it Spitzer}/GLIMPSE $5.8\,\mu$m image of the same area. The red circle shows the O7\,II star a805, which we identify as the ionising source, and its mid-infrared counterpart (G025.2465+00.3011). The blue square marks the position of A514 and its bright GLIMPSE counterpart (G025.2977+00.3105), for reference.  \label{spitzer}}
\end{figure*}

\begin{figure}
   \centering
   \resizebox{\columnwidth}{!}{\includegraphics[angle=-90]{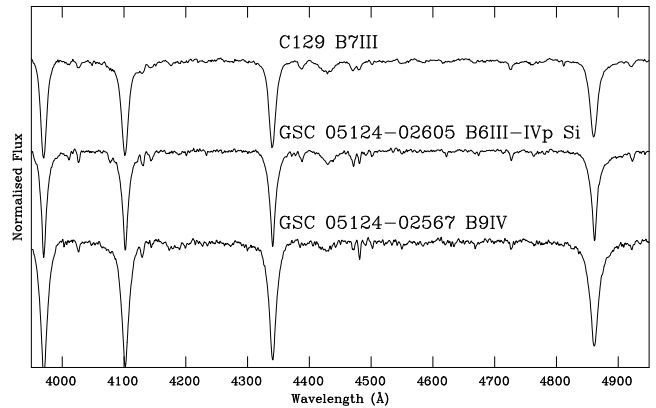}}
   \caption{Classification spectra of three stars that can only be foreground to the Alicante~6 cluster. GSC~05124-02567 and GSC~05124-02605 are included in the work of \citet{roslund1963} and fall outside our photometric fields. C129 is the object marked B7\,III in Fig.~\ref{mapmembers}.\label{fig:oldstars}
   }
    \end{figure}

Figure~\ref{spitzer} shows the DSS2 red image of the area with brightest nebulosity. We find a very high number of early type stars immersed in this nebulosity. We have spectra for six of these objects. Even though the spectra are noisy, because the stars are heavily reddened and thus very faint in $B$, we can confidently assign spectral types between B0\,V and B1\,V to all of them. In total, we identify $\approx20$ cluster members in this area of bright
nebulosity. Star a805 with spectral type O7\,II is the most luminous object that we have identified and likely the main 
source of ionisation of the bright nebulosity. It lies approximately $2\farcm5$ from the main cluster. The brightest nebulosity seems to form an arc around this position, again suggesting that this is the main ionising source. Indeed the geometrical configuration resembles other regions where sequential or triggered star formation has been claimed \citep[e.g.,][]{zavagno06,zavagno07}. The fact that a805 is already close to being a supergiant suggests that an age difference could exist, favouring the idea that the emerging cluster has been triggered by the luminous star. An exploration of the 2MASS CMDs for the whole area, using criteria to select early-type stars \citep[cf.][]{cp05,ns07}, indicates that a805 is the brightest early-type star in $K_{{\rm S}}$ over the whole area. Apart from the three objects seen in the optical, we cannot find any early-type star sufficiently bright in $K_{{\rm S}}$ to be an O-type star, unless it is very heavily ($A_{V}\ga20$~mag) embedded. 

In Fig.~\ref{spitzer}, we also show the Spitzer/GLIMPSE $5.8\,\mu$m image of the same area. Though most of the optically visible stars have GLIMPSE counterparts, they are generally faint in the mid-infrared. There are, however, several other bright point sources in the near and mid-infrared images with faint or no counterparts in the optical images (for example, G025.3016+00.2862 and G025.2978+00.2791 in the vicinity of A507). 
The area containing the \ion{H}{ii} nebulosity and the optically visible OB stars appears as a hole in the $5.8\,\mu$m images. The cavity is surrounded by a rim of emitting material, prominent at $5.8$ and $8.0\,\mu$m, which is listed as N37 in the catalogue of bubble-like structures by \citet{churchwell2006} and was imaged by \citet{dehar10}. This suggests that the young stars have carved a cavity in the molecular cloud. Some of the bright $5.8\,\mu$m sources could be still embedded OB stars. The {\it Spitzer} image detects a small cluster of mid-infrared sources around the position of the star A514, which is the only optically identified member coincident with a bright {\it Spitzer} source (G025.2977+00.3105, with $8.0\,\mu$m magnitude $4.46\pm0.04$). Indeed, the stars A514 and A515 have a value of $M_{V}$ that is slightly too high (by $\sim1$~mag) for their spectral types. Their {\it Spitzer} counterparts suggest that they may be still deeply embedded or, perhaps more likely, be unresolved small groups of stars.

It is difficult to estimate the age of the cluster because of the lack of evolved stars. The O7\,II star is likely to have an age 2--3~Myr, because of its spectral type \citep[cf.][]{negueruela08}. If the triggered formation scenario is correct, the age of the emerging cluster should be lower. The field is too complex to attempt any further discussion of the embedded population without a deep infrared study.

\subsection{Different population along the line of sight}

Previous studies concentrated on the optically brightest stars in the area. \citet{vogt75} suggested that star a802 could be the ionising source of the \ion{H}{ii} region. We find a802 to have spectral type B1\,II. Therefore, it is a massive, luminous star, but probably too cool to contribute significantly to the ionisation of the \ion{H}{ii} region, when compared to the nearby O-type star. Moreover, a802 is less reddened than any of the O-type stars.

Star a803 ([R63]~30 = GSC~05124-02549), which is also in \citet{roslund1963}, has spectral type B2\,III-IV. It has a reddening of $E(B-V)=0.95$, much lower than the stars associated with the nebulosity. Star a801 ( = [R63]~36; B1\,V) has a similar $E(B-V)=0.94$. This lower reddening might be caused by these objects lying far away from the molecular cloud, but the spectral types B1\,II and B2\,III-IV are not expected in a cluster emerging from its parental cloud, such as Alicante~6 seems to be, nor to be compatible with the expected age of an O7\,II star. In spite of this, all three stars closely follow the cluster ZAMS and have spectroscopic distance moduli compatible with the cluster distance. They lie very close in the sky to the emerging cluster Alicante~6. Indeed a803 lies only $1 \arcmin$ away from a805, the O7\,II star that we identify as the main ionisation source of RCW~173, for which $E(B-V)=2.15$. 

This mixture of stars with very different reddenings and spectral types corresponding to ages, but sharing the same distance, suggests that we are seeing the projection of several populations of early-type stars at approximately the same distance, but having formed over a period longer than $\ga10\:$Myr. These stars could be part of an OB association related to Alicante~6 and projected along the line of sight. Even when exactly projected along the line of sight, the typical size of an OB association (50--100~pc) cannot be distinguished via photometric distances at $d=3$~kpc. They could also be part of an unrelated population of early-type stars at a slightly lower distance. Examples of star-forming complexes producing massive stars over several Myr are not uncommon \citep[e.g.,][]{cp04,clark2009b}, though age spreads as large as 10~Myr are infrequent.

To the south of the cluster, we find two other O-type stars, b201 and C81 (both O7\,V and rather heavily reddened) surrounded by many other stars at the same distance, some of them with lower reddenings, e.g., C84 (B1\,V) with $E(B-V)=1.12$. It does not seem possible to decide whether the stars with moderate reddening ($E(B-V)\la1.2$) represent a different population from the stars with higher reddening. We again may speculate that objects with low reddening are in the foreground (where foreground here indicates lying in front of the cluster, though likely belonging to the same OB association) while objects with high reddenings are directly associated with the \ion{H}{ii} region. Star C81 is far too bright for its spectral type (more than 1~mag). Its spectrum shows evidence of at least one other early-type star blended with the O7\,V object, as its lines look diluted. We do not rule out membership for this star.

In addition, three of the stars from \citet{roslund1963}, GSC~05124-02605, GSC~05124-02567, and  GSC~05124-02627 (none of which are inside our photometric field), have much lower distance moduli. Star GSC~05124-02567 (B9\,IV) has a very low distance modulus and lies only $\sim600$~pc away. Stars GSC~05124-02605 (B6\,III-IVp Si) and GSC~05124-02627 (B3\,III) have $DM\approx11.5$ ($d\approx2\:$kpc). Another star clearly unrelated to Alicante~6, C129 (B7\,III) has a similar distance modulus. All these objects have $E(B-V)\approx1.0$. 

Our results show a great similarity to those obtained by \citet{turner1980}, who analysed photometry and spectra for a number of luminous stars in the region surrounding the open cluster Trumpler~35 (towards $l=28\degr$). He found two distinct groupings of stars placed at two different distances: a moderately young population, containing B-type giants reddened by $E(B-V)\approx1.0$ at $\sim2$~kpc (in the Sagittarius arm) and a very young population (consisting of O-type stars and bright supergiants) with similar or slightly higher colour excesses at a distance of $3.5$~kpc. Along our line of sight, we also identify B-type giants at $\sim2$~kpc, which should be tracing the Sagittarius arm. Over the whole $l=24\degr-29\degr$ range, the Sagittarius arm does not seem to contain any star-forming region, and only moderately young clusters. NGC~6664 has an estimated age of $\sim50$~Myr \citep{schmidt82}, while the age of Trumpler~35 has been inferred tobe between 30 and 50~Myr \citep[and references therein]{turner1980}. Other clusters in this direction, such as Basel~1 or NGC~6704, have distances of $d\sim1.5$~kpc and older ages \citep{delgado97}.

In this direction, the line of sight crosses the Sagittarius arm and then intersects the Scutum-Crux arm almost tangentially (for instance, \citealt{green11} place the origin of the arm at $l=26\degr$, based on the concentration of methanol masers, while the distribution of \ion{H}{ii} regions indicates that the tangent point is around $l=31\degr$; \citealt{bania2010}). Our data show a population of OB stars at $d\sim 3$~kpc with essentially the same $E(B-V)$ as the objects in the Sagittarius arm. Towards $l=28-29\degr$, \citet{turner1980} also finds a very young population with the same $E(B-V)$ as Sagittarius arm stars in that direction. We identify these populations as the near intersection with the Scutum-Crux arm. Along our line of sight, we find the molecular cloud associated with RCW~173 immediately behind. Obscuration by this cloud results in the much higher reddenings observed in the members of Alicante~6. For a nearby sightline at ($l=24\fdg6$, $b=+0\fdg4$), \citet{negueruela10a} find an abrupt increase in the infrared extinction between $d=3.0$ and $3.4$~kpc, in excellent agreement with the presence of obscuring clouds detected here.

\subsection{Connection with the W42 complex}

Using a Fabry-Perot spectrometer, \citet{fich90} measure the velocity $v_{{\rm LSR}}=36.5\pm0.3\:{\rm km}\,{\rm s}^{-1}$ for the H$\alpha$ emission in Sh2-60 (RCW~173). On the other hand, \citet{blitz82} measure a velocity $v_{{\rm LSR}}=43.8\pm2.1\:{\rm km}\,{\rm s}^{-1}$ for the CO in the associated molecular cloud. The difference between these two values is typical of \ion{H}{ii} regions \citep{fich90}. These velocities are perfectly compatible with a distance of 3.0~kpc \citep[see the Galactic rotation curve towards $l=25\fdg3$ in][]{davies2008}, giving very strong support to our photometric distance\fnmsep\footnote{Note that \citet{dehar10} adopt the far dynamical distance for N37, $d=12.6$~kpc. Such a high distance is completely incompatible with the detection in $U$ of the associated stellar population.}. Other nearby \ion{H}{ii} regions, such as Sh2-59 ($l=24\fdg5$, $b=-0\fdg2$) or Sh2-61 ($l=26\fdg4$, $b=+1\fdg8$), have very similar dynamical distances, supporting the idea that we are seeing the edge of the Scutum-Crux arm, delineated by molecular clouds.

Recently, \citet{rm10} considered the possibility of a large star-forming complex over the $l=23\degr-26\degr$ range, associated with the W42 star-forming region. \citet{blum00} found an embedded cluster within W42, containing at least one O-type star, at only $\sim30\arcmin$ from RCW~17. However, \citet{kolpak03} found dynamical distances of 4.0 and 4.5~kpc for the G25.38$-$0.181 and G25.38$-$0.181 \ion{H}{ii} complexes, which seem to be the core of this putative W42 star-forming complex, while several other nearby regions have $v_{{\rm LSR}}$ in the $90-110\:{\rm km}\,{\rm s}^{-1}$ range, corresponding to dynamical distances $d\sim 6-7$~kpc \citep{rm10}. These very different values favour the idea that we are looking along the Scutum-Crux over a distance of several kpc, rather than supporting a single star-forming complex. 

\section{Conclusions}

We have analysed photometry and spectroscopy for stars in the direction of the \ion{H}{ii} region RCW~173 (Sh2-60). We have found that most stars in the field are reddened B-type stars, according to their $Q$ values. These B stars have a broad range of reddenings and are spread over a range of distances. In particular, we have discovered:

\begin{enumerate}
\item A heavily reddened ($E(B-V)\ga1.5$) population, including three O-type stars. About 20 early-type stars are concentrated in a small region $\sim3\arcmin$ across, coincident with the brightest nebulosity. Spectra for six of these objects show all of them to be B0--1\,V stars. We identify this region (centred around RA 18h 36m 23s, Dec $-$06 39 40) as an emerging young open cluster, which we call Alicante~6. The cluster is associated with a number of strong mid-infrared point sources visible in {\it Spitzer}/GLIMPSE images and lies inside a bright bubble-like rim of emission, visible at $5.8$ and $8.0\,\mu$m, which seems to be an evacuated cavity. 

\item Star a805, located $\sim2.5\arcmin$ SW of Alicante~6, has spectral type O7\,II, and is likely the main source of ionisation in the area. It may have triggered the formation of the cluster.

\item A less reddened ($E(B-V)\approx0.9-1.2$) population of early-type stars is spatially coincident with the more obscured objects, which include at least two early-B giants and seem to be located at the same distance as Alicante~6. In particular, star a802 (GSC~05124-02611), previously suggested as the ionising source, has spectral type B1\,II and is the second brightest star in the area. This population, which must be several Myr old, cannot be separated from the very young population with the data available. We speculate that it represents an earlier generation in the same area, which could be identified with an OB association.

\item A small number of B giants with distances compatible with a location in the Sagittarius arm ($d\sim2$~kpc).

\end{enumerate}

These different populations testify to the complexity of this line of sight. We interpret the bulk of stars at $d\approx3.0$~kpc as an OB association at the near intersection of the Scutum-Crux arm, lying just in front of a complex of molecular clouds. The very young cluster Alicante~6 (with a likely age $\la2$~Myr) is emerging from its parental cloud, its formation having perhaps been triggered by the nearby O7\,II star. Infrared and radio observations reveal a wealth of star-forming sites at larger distances, consistent with two geometrical configurations. The sightline may cross the Scutum-Crux arm (with a thickness of $\sim1$~kpc) and then meet the tip of the Galactic bar around $d\approx6$~kpc, or, alternatively, run along the length of the Scutum-Crux arm (or, equivalently, a Molecular Ring) for $\sim3$~kpc. Further study of the stellar populations in this and neighbouring directions is needed to clarify this issue. In any event, given the complexity of lines of sight in this direction, any attempt at quantifying the total mass and time-averaged formation rate in this putative complex of clusters of red supergiants must take into account the need to distinguish star-forming regions at different distances.

\begin{acknowledgements}

We thank Simon Clark and Carlos Gonz\'alez-Fern\'andez for useful comments and  the anonymous referee for helpful suggestions.

Based on observations made with the Nordic Optical Telescope, operated
on the island of La Palma jointly by Denmark, Finland, Iceland,
Norway, and Sweden, in the Spanish Observatorio del Roque de los
Muchachos of the Instituto de Astrofisica de Canarias.  

The data presented here have been taken using ALFOSC, which is owned by the Instituto de Astrofisica de Andalucia (IAA) and operated at the Nordic Optical Telescope under agreement between IAA and the NBIfAFG of the Astronomical Observatory of Copenhagen

This research is partially supported by the MEC under grants AYA2008-06166-C03-03, AYA2010-21697-C05-05 and CSD2006-70. 
This research has made use of the Simbad database, operated at CDS, Strasbourg (France). This publication
makes use of data products from the Two Micron All Sky Survey, which is a joint project of the University
of Massachusetts and the Infrared Processing and Analysis Center/California Institute of Technology, funded
by the National Aeronautics and Space Administration and the National
Science Foundation. 
\end{acknowledgements}

\clearpage

\begin{sidewaystable}
\caption{$(X,Y)$ position on the map A of stars with photometry in the field. Photometry for stars in frame A. 2MASS (or GLIMPSE) identification for these stars and their coordinates.\label{t2}}
\begin{tabular}{lllllllllllll}
\hline
\hline
 Name&X (Pixels)&Y (Pixels)&$V$&$\sigma_{V}$&$(B-V)$&$\sigma_{(B-V)}$&$(U-B)$&$\sigma_{(U-B)}$&$N$
&RA (J2000)&DEC (J2000)& Name(2MASS)\\
\hline
A6&1432.826&27.839      &18.680	&0.009	&1.569	&0.015	&0.975	&0.018	&4&18 36 15.892	&-06 44 40.00	&18361589-0644400\\
\hline
\end{tabular}
\end{sidewaystable}

\begin{sidewaystable}
\caption{$(X,Y)$ position on the map B of stars with photometry in the field. Photometry for stars in frame B. 2MASS (or GLIMPSE) identification for these stars and their coordinates.\label{t3}}
\begin{tabular}{lllllllllllll}
\hline\hline
 Name&X (Pixels)&Y (Pixels)&$V$&$\sigma_{V}$&$(B-V)$&$\sigma_{(B-V)}$&$(U-B)$&$\sigma_{(U-B)}$&$N$
&RA (J2000)&DEC (J2000)& Name(2MASS)\\
\hline
B1&1620.152	&	15.78	&	15.085	&	0.011	&	2.104	&	0.013	&	2.053	&	0.025	&	1	&	18 36 09.902	&	-06 50 29.84	&	18360990-0650298\\
\hline
\end{tabular}
\end{sidewaystable}

\begin{sidewaystable}
\caption{$(X,Y)$ position on the map C of stars with photometry in the field. Photometry for stars in frame B. 2MASS (or GLIMPSE) identification for these stars and their coordinates.\label{t4}}
\begin{tabular}{lllllllllllll}
\hline\hline
 Name&X (Pixels)&Y (Pixels)&$V$&$\sigma_{V}$&$(B-V)$&$\sigma_{(B-V)}$&$(U-B)$&$\sigma_{(U-B)}$&$N$
&RA (J2000)&DEC (J2000)& Name(2MASS)\\
\hline
C2&1547.672	&	12.715	&	17.745	&	0.010	&	1.447	&	0.064	&	0.799	&	0.036	&	3	&	18 36 21.044	&	-06 50 04.82	&	18362104-0650048\\
\hline
\end{tabular}
\end{sidewaystable}

\begin{table}
\caption{Derived parameters for stars in Alicante 6 and the surrounding association.\label{t7}}
\begin{tabular}{lllllll}
\hline
\hline
 $Number$&$E(B-V)$&$\sigma_{E(B-V)}$&$A_{V}$&$V_{0}$&$(B-V)_{0}$&$M_{V}$\\
\hline
A12&1.598&0.019&4.954&13.121&-0.143&0.721\\				
\hline
\end{tabular}
\end{table}
												

\begin{thebibliography}{}

\bibitem[Bally et al.(2010)]{bally10} Bally, J., Anderson, L.D., Battersby, C., et al. 2010, A\&A, 518, L90

\bibitem[Bania et al.(2010)]{bania2010} Bania, T.M., Anderson, L.D., Balser, D.S., \& Rood, R.T. 2010, ApJL 718, L106

\bibitem[Benjamin et al.(2003)]{glimpse} Benjamin, R.A., Churchwell, E., Babler, B.L., et al. 2003, PASP, 115, 953

\bibitem[Blitz et al.(1982)]{blitz82} Blitz, L., Fich, M., \& Stark, A.A. 1982, \apjs, 49, 183

\bibitem[Blum et al.(2000)]{blum00} Blum, R.D., Conti, P.S., \& Damineli, A. 2000, AJ, 119, 1860 

\bibitem[Clark \& Porter(2004)]{cp04} Clark, J.S., \& Porter, J.M. 2004, A\&A, 427, 839

\bibitem[Clark et al.(2009a)]{clark2009a} Clark, J.S., Negueruela, I., Davies, B., et al. 2009a, A\&A, 498, 109

\bibitem[Clark et al.(2009b)]{clark2009b} Clark, J.S., Davies, B., Najarro, F., et al. 2009b, A\&A, 498, 109


\bibitem[Churchwell et al.(2006)]{churchwell2006} Churchwell, E., Povich, M. S., Allen, D., et al. 2006, ApJ, 649, 759 

\bibitem[Comer\'on \& Pasquali(2005)]{cp05}Comer\'on, F., \& Pasquali,
  A. 2005, A\&A, 430, 541

\bibitem[Delgado et al.(1997)]{delgado97} Delgado, A.J., Alfaro, E.J., \& Cabrera-Cano, J. 1997, AJ, 113, 713

\bibitem[Draper et al.(2000)]{Draper2000} Draper, P.W., Taylor, M., \& Allan, A. 2000, Starlink User Note 139.12, R.A.L.

\bibitem[Davies et al.(2007)]{davies2007} Davies, B., Figer, D.F., Kudritzki, R.P., et al. 2007, ApJ, 671, 781 (D07)

\bibitem[Davies et al.(2008)]{davies2008} Davies, B., Figer, D.F., Law, C.J., et al. 2008, ApJ, 676, 1016

\bibitem[Deharveng et al.(2010)]{dehar10} Deharveng, L., Schuller, F., Anderson, L.D., et al. 2010, A\&A 523, A6

\bibitem[Fitzgerald(1970)]{fitzgerald1970} FitzGerald, M.P. 1970, A\&A, 4, 234

\bibitem[Fich et al.(1990)]{fich90} Fich, M., Treffers, R.R., \& Dahl, G.P. 1990, AJ, 99, 622

\bibitem[Garz\'on et al.(1997)]{garzon1997} Garz\'on, F., L\'opez Corredoira, M., Hammersley P., et al. 1997, ApJ, 491, L31

\bibitem[Green et al.(2011)]{green11} Green, J.A., Caswell, J.L., McClure-Griffiths, N.M., et al. 2011, ApJ, in press ({\tt arXiv:1103.3913})

\bibitem[Howarth et al.(1998)]{Howarth1998} Howarth, I., Murray, J., Mills, D., \& Berry, D.S. 1998, Starlink User Note 50.21, R.A.L.


\bibitem[Johnson \& Morgan(1952)]{johnson1952} Johnson, H.L., \& Morgan, W.W. 1952, ApJ, 117, 313

\bibitem[Kolpak et al.(2003)]{kolpak03} Kolpak, M.A., Jackson, J.M., Bania, T.M., et al. 2003, ApJ, 582, 756

\bibitem[Lahulla(1985)]{lahulla85} Lahulla, J. F. 1985, A\&AS, 61, 537

\bibitem[Landolt(1992)]{landolt} Landolt, A.U. 1992, AJ, 104, 340

\bibitem[Ma\'{i}z-Apell\'aniz(2004)]{jesus2004} Ma\'{i}z-Apell\'aniz, J. 2004, PASP, 116, 859


\bibitem[Morton \& Adams (1968)]{morton68} Morton, D.C., \& Adams, T.F. 1968, ApJ, 151, 611

\bibitem[Negueruela \& Schurch(2007)]{ns07} Negueruela, I., \& Schurch,
  M.P.E. 2007, A\&A, 461, 431

\bibitem[Negueruela et al.(2008)]{negueruela08} Negueruela, I., Marco, A., Herrero, A. \&  Clark, J. S. 2008, A\&A, 487, 575

\bibitem[Negueruela et al.(2010)]{negueruela10a} Negueruela, I., G\'onzalez-Fern\'andez, C., Marco, A., Clark, J.S. \& Mart\'{\i}nez-N\'u\~{n}ez, S. 2010, A\&A, 513, A74

\bibitem[Negueruela et al.(2011)]{negueruela11} Negueruela, I., G\'onzalez-Fern\'andez, C., Marco, A.,  \& Clark, J.S. 2011, A\&A, 528, A59 

\bibitem[Nguyen Luong et al.(2011)]{nguyen11} Nguyen Luong, Q., Motte, F., Schuller, F., et al. 2011, A\&A, 529, A41

\bibitem[Rahman \& Murray(2010)]{rm10} Rahman, M., \& Murray, N. 2010, ApJ, 719, 1104

\bibitem[Rodgers et al.(1960)]{rodgers60} Rodgers, A. W., Campbell, C. T., Whiteoak, J. B. 1960, MNRAS, 121, 103

\bibitem[Roslund(1963)]{roslund1963} Roslund, C. 1963, Arkiv f\"or Astronomi, 3, 97

\bibitem[Schmidt(1982)]{schmidt82} Schmidt, E.G. 1982, AJ, 87, 1197

\bibitem[Schmidt-Kaler(1982)]{skaler82} Schmidt-Kaler, T. 1982, Landolt-B\"ornstein, N.S. VI, 2b

\bibitem[Sharpless(1959)]{sharpless59} Sharpless, S. 1959, ApJS, 4, 257

\bibitem[Shortridge et al.(1997)]{Shortridge1997} Shortridge, K., Meyerdicks, H., Currie, M., et al. 1997, Starlink User Note 86.15, R.A.L.


\bibitem[Skrutskie et al.(2006)]{Skrutskie2006} Skrutskie, M.F., Cutri, R.M., \& Stiening, R. 2006, AJ, 131, 1163

\bibitem[Stetson(1987)]{stetson1987} Stetson, P. B. 1987, PASP, 99, 191

\bibitem[Turner(1980)]{turner1980} Turner, D. 1980, ApJ, 240, 137

\bibitem[Vogt \& Moffat (1975)]{vogt75} Vogt, N., Moffat, A. F. J. 1975, A\&A, 45, 405

\bibitem[Walborn \& Fitzpatrick (1990)]{walborn1990} Walborn, N.R. \& Fitzpatrick, E.L. 1990, PASP, 102, 379

\bibitem[Zavagno et al.(2006)]{zavagno06} Zavagno, A.,  Deharveng, L, Comer\'on, F., et al. 2006, A\&A, 446, 171

\bibitem[Zavagno et al.(2007)]{zavagno07} Zavagno, A., Pomar\`es, M., Deharveng, L, et al. 2007, A\&A, 472, 835


\end{thebibliography}
\end{document}